\documentstyle[emulateapj,psfig]{article}

\def\ale{\mathrel{\hbox{\rlap{\hbox{\lower4pt\hbox{$\sim$}}}\hbox{$<$}}}}
\def\age{\mathrel{\hbox{\rlap{\hbox{\lower4pt\hbox{$\sim$}}}\hbox{$>$}}}}

\def\gsim{\mathrel{\hbox{\rlap{\lower.55ex \hbox {$\sim$}}
                   \kern-.3em \raise.4ex \hbox{$>$}}}}
\def\lsim{\mathrel{\hbox{\rlap{\lower.55ex \hbox {$\sim$}}
                   \kern-.3em \raise.4ex \hbox{$<$}}}}

\def\etal{et al.}

\def\kms{km s$^{-1}$}

\lefthead{Djorgovski \etal~ }
\righthead{Spectroscopy of GRB 980703}

\begin{document}

\title{Spectroscopy of the Host Galaxy of the Gamma--Ray Burst 980703
\footnotemark}
\footnotetext{
Based in part on observations obtained at the W. K. Keck Observatory, which
is operated by the California Association for Research in Astronomy, a
scientific partnership among the California Institute of Technology, the
University of California, and the National Aeronautics and Space Administration.
}

\author{S. G. Djorgovski$^2$, S. R. Kulkarni$^2$, J. S. Bloom$^2$,}

\author{R. Goodrich$^3$, D. A. Frail$^4$, L. Piro$^5$, and E. Palazzi$^6$}

\bigskip 
\bigskip 

\affil{$^2$ Palomar Observatory, California Institute of Technology,
            Pasadena, CA 91125, USA}

\affil{$^3$ W. M. Keck Observatory, Kamuela, HI 96743, USA}

\affil{$^4$ National Radio Astronomy Observatory, Socorro, NM 87801, USA}

\affil{$^5$ Istituto di Astrofisica Spaziale, CNR, Via Fermi 21, 
            00044 Frascati, Italy} 

\affil{$^6$ Istituto Tecnologie e Studio Radiazioni Extraterrestri, CNR, 
            40129 Bologna, Italy}

\begin{abstract}

We present spectroscopic observations of the host galaxy of the $\gamma$-ray
burst (GRB) 980703.  Several emission and absorption features are detected,
making the redshift, $z = 0.966$, completely unambiguous.  This is only the
third known redshift for a GRB host.  The implied isotropic $\gamma$-ray energy
release from the burst is in excess of $10^{53}$ erg, for a reasonable choice
of cosmological parameters.  The spectroscopic properties of the host galaxy
are typical for a star formation powered object.  Using the observed value of
the Balmer decrement, we derived the extinction in the galaxy's restframe, $A_V
\approx 0.3 \pm 0.3$ mag.  Using three different star formation rate
indicators, we estimate $SFR \approx 10 ~M_\odot ~{\rm yr}^{-1}$, or higher,
depending on the extinction, with a lower limit of 
$SFR > 7 ~M_\odot ~{\rm yr}^{-1}$.  This is the highest value of the
star formation rate measured for a GRB galaxy so far, and it gives some
support to the idea that GRBs are closely related to massive star formation.

\end{abstract}


\section{Introduction}

Studies of the cosmic $\gamma$--ray bursts (GRBs; \cite{kle73})
are currently one of the most active fields of research in astronomy.  
Much of the recent progress in this area was enabled by the discovery of 
long-lived x-ray afterglows of GRBs (e.g., \cite{cos97}),
made possible by the BeppoSAX satellite (\cite{boe97}).  This, in turn, 
enabled the discovery of optical 
(e.g., \cite{jvan97}, \cite{bond97}, \cite{djo97}, \cite{hal98}, 
\cite{jau98}, \cite{hjo98}, etc.), 
and radio 
(e.g., \cite{fra97}, \cite{tay98})
afterglows and their subsequent studies, including the determination of their
distances and various physical parameters.

The key measurement is the determination of redshifts.  This has demonstrated
convincingly the cosmological nature of GRBs (cf. \cite{pac95}), and determined
the energetics scale of this fascinating phenomenon. 
To date, redshifts have been obtained for only two GRB host galaxies: 
$z = 0.835$ for GRB 970508 (\cite{met97}, \cite{bloom98a}), 
and $z = 3.428$ for GRB 971214 (\cite{kul98}).  
The two cases have implied energy releases which differ by more than an order
of magnitude 
(\cite{djo97}, \cite{ram98}, \cite{kul98}).
Obviously, there is a pressing need to increase the size of this sample. 
Moreover, the physical properties of GRB host galaxies, especially the star
formation rates, provide essential clues about the possible physcial origins 
of the GRB phenomenon. 

Following the discovery of GRB 980703 by the RXTE satellite (\cite{lev98}),
a variable x-ray source was localised in its error-box by the BeppoSAX
satellite (\cite{gal98a}, \cite{gal98b}).  
At the same time, radio observations from the VLA and optical observations from
the Keck and Palomar lead to the discovery of a radio and optical transient
(OT) within the error-circle of the fading x-ray source 
(\cite{fra98a}, \cite{fra98b}, \cite{bloom98b}). 
\cite{bloom98c} report in detail on these observations.
This was subsequently confirmed by \cite{zapos98} and \cite{castro98}.
In this Letter we report on the spectroscopy of the host galaxy of GRB 980703.
Our preliminalry results have been reported in \cite{djo98}.  

\section{Observations and Data Reductions}

Spectra of the host galaxy were obtained using the Low Resolution Imaging
Spectrograph (LRIS; \cite{Oke95})
mounted at the Cassegrain focus of the Keck II 10-m telescope, on UT 1998 July
07.6 and 19.6, in photometric conditions, and with sub-arcsecond seeing.  There 
was some moonlight on July 19.  On July 7 we used a 300 lines mm$^{-1}$ grating
giving a typical resolution of $FWHM \approx 12$ \AA, and a usable wavelength
range from approximately 3800 to 8800 \AA.  The total exposure was 1 hour. 
Exposures of the spectroscopic standard BD +17$^\circ$4708 (\cite{OkeGunn})
were used for flux calibration.  On July 19 we used an 831 lines mm$^{-1}$
grating giving a typical resolution of $FWHM \approx 4$ \AA, and a usable
wavelength range from approximately 7900 to 9800 \AA.  The total exposure was
0.5 hours.  Exposures of the spectroscopic standard BD +28$^\circ$4211 
(\cite{mas88}) 
were used for flux calibration.  

Exposures of arc lamps obtained right after the target observations were used
for the wavelength calibration, with a resulting r.m.s. uncertainty of about
0.3 \AA\ for the July 7 data, and 0.1 \AA\ for the July 19 data, and with the
possible systematic errors of the same order, due to the instrument flexure. 
All observations were obtained with the slit position angle close to the
parallactic angle, so that the atmospheric refraction effects were minimized. 
The estimated flux zero-point calibration is less than 10\%.
We tested the flux calibration consistency of the two spectra by comparing the 
line fluxes of two emission lines detected in both observations (identified as
H$\delta$ and H$\gamma$; see below); these fluxes should remain constant, while
the continuum intensity was expected to change, due to the fading of the OT. 
The line fluxes agree to better than 10\%. 

The resulting spectra are shown in Figs. 1 and 2.  Several strong emission 
lines are observed, as listed in Table 1.  The estimated errors of the
observed line fluxes are less than 5\% internal (due to the random noise and
the uncertainties in the continuum level determination), plus about 10\%
external, due to the uncertainties in the flux calibration, including 
aperture corrections and other possible systematics.
In addition, several absorption features are seen in the blue part of the
spectrum, as shown in Fig. 3 and listed in Table 2.  The mean emission-line
redshift is 
$z_{em} = 0.9662 \pm 0.0002$, 
and the mean absorption redshift is
$z_{abs} = 0.9656 \pm 0.0006$.
The corresponding restframe velocity difference is $\Delta V = (90 \pm 110)$
\kms, which is consistent both with zero and with velocity fields expected
in normal galaxies.

The observed continuum flux contains light from both the host galaxy and the OT.
Our spectroscopic measurements give magnitudes 
$B \approx 23.3$,
$V \approx 22.65$,
$R \approx 21.9$, and
$I \approx 21.3$ mag for the July 7.6 data, and
$I \approx 22.0$ mag for the July 19.6 data, uncertain by about 0.2 mag.
These are in a good agreement with the magnitudes estimated for these epochs
from our direct imaging (\cite{bloom98c}).
We note that the OT accounted for about one-third of the observed flux in
the $R$ and $I$ bands on July 7, and the observed equivalent widths of emission
lines should be corrected accordingly, i.e., relative to the galaxy's light
they should be increased by about 50\%.  The contribution of the OT on July 19
was negligible.

We estimate the foreground Galactic reddening in this direction from the maps
by \cite{Sch98} to be $E_{B-V} = 0.061$ mag.  Using the Galactic extinction
curve from \cite{Car89}, we obtain extinction values of
0.251, 0.188, 0.141, 0.090, 0.053, 0.036, and 0.021 mag in the $BVRIJHK$ bands,
respectively.  Table 1 also lists the extinction-corrected emission line fluxes.

\section{Discussion}

This is only the third known redshift for a host galaxy of a GRB.
Assuming a standard Friedman model cosmology with $H_0 = 65$ \kms\ Mpc$^{-1}$
and $\Omega_0 = 0.2$, for $z = 0.9662$ we derive a luminosity distance 
$d_L  = 1.921 \times 10^{28}$ cm.  

The observed $\gamma$-ray fluence of this burst was 
$(4.59 \pm 0.42) \times 10^{-5}$ erg cm$^{-2}$ 
(\cite{kip98}; cf. also \cite{amati}).
The corresponding isotropic energy release for our assumed cosmology is
$(1.1 \pm 0.1) \times 10^{53}$ erg, making it nearly as luminous as
GRB 971214 (\cite{kul98}, \cite{ram98}), and thus showing that GRB 971214
was not peculiar in this regard.

The low-ionization emission-line spectrum of the host galaxy of GRB 980703 is
typical of an actively star-forming galaxy.  There is no sign of an active 
nucleus, given the absence of any high-ionization lines, and the relative
weakness of the [O III] 4959 line compared to H$\beta$. 

The observed equivalent width of the [O II] 3727 line on July 7 was $(91 \pm
5)$ \AA, or about 46 \AA\ in the galaxy's restframe.  Correcting for the
estimated continuum contribution from the OT would increase these numbers by
about 50\%, placing this object at the high end of the distribution for the
typical field galaxies at comparable magnitudes and redshifts
(\cite{Hogg98}). 

Given that we observe three lines of the Balmer series, we can make an
estimate of the effective extinction in the galaxy's star forming regions.
We observe line ratios 
H$\gamma$/H$\beta$ = $0.46 \pm 0.01$
and 
H$\delta$/H$\beta$ = $0.23 \pm 0.01$ 
(note that all systematic errors in flux calibration cancel for these ratios).
We assume that the fractional correction due to the fill-in of the stellar
absorption lines in the spectrum is the same for all three Balmer lines.
The corresponding ranges for the standard Case B recombination (\cite{oster})
for a broad range of temperatures and electron densities are
H$\gamma$/H$\beta$ = $0.469 \pm 0.009$
and 
H$\delta$/H$\beta$ = $0.229 \pm 0.004$.
The ratios of the observed and theoretical values are thus
$0.98 \pm 0.04$ and $0.89 \pm 0.06$, respectively, indicating that some
extinction may be present at the source, which is not surprising.
Using the Galactic extinction curve (\cite{Car89}) and the foreground screen
approximation, the implied restframe extinction values are
$A_V \approx 0.125 \pm 0.25$ mag from the H$\gamma$/H$\beta$ ratio, and
$A_V \approx 0.47 \pm 0.28$ mag from the H$\delta$/H$\beta$ ratio, with the
weighted average $A_V \approx 0.3 \pm 0.3$ mag (in the galaxy's restframe).

We can estimate the star formation rate in this object in three different ways:

The implied [O II] line luminosity, corrected for the Galactic foreground, but
not for any restframe extinction, is 
$L_{3727} = (1.41 \pm 0.15) \times 10^{42}$ erg s$^{-1}$.
Assuming that the line is powered by star formation alone, the conversion from
\cite{Ken98},
we estimate the star formation rate 
$SFR \approx (20 \pm 7) ~M_\odot ~{\rm yr}^{-1}$.
Adopting our estimated restframe extinction of $A_V \approx 0.3$ mag would
increase these numbers by 53\%, i.e., to 
$SFR \approx 30 ~M_\odot ~{\rm yr}^{-1}$.

Our observed H$\beta$ line flux, corrected only for the Galactic foreground
extinction, is
$f_{H\beta} = (6.1 \pm 0.6) \times 10^{-17}$ erg cm$^{-2}$ s$^{-1}$.
Using the theoretical, Case B recombination value of
H$\alpha$/H$\beta$ = $2.85 \pm 0.1$
(\cite{oster}),
we derive the predicted H$\alpha$ flux
$f_{H\alpha} = (1.74 \pm 0.19) \times 10^{-16}$ erg cm$^{-2}$ s$^{-1}$.
The implied line luminosity (again, not corrected for any restframe extinction)
is 
$L_{6563} = (8.1 \pm 0.9) \times 10^{41}$ erg s$^{-1}$.
Using the conversion from \cite{Ken98},
we estimate the star formation rate 
$SFR \approx (6.4 \pm 0.7) ~M_\odot ~{\rm yr}^{-1}$.
Adopting restframe extinction of $A_V \approx 0.3$ mag would
increase these numbers by 25\%, i.e., to 
$SFR \approx 8 ~M_\odot ~{\rm yr}^{-1}$.
Despite the extrapolation from H$\beta$ to H$\alpha$, this is probably a more
reliable estimate than the star formation rate derived from the
density-sensitive [O II] line. 

However, the observed Balmer line luminosities are likely to be underestimated,
due to the underlying stellar absorption features at the same wavelengths.
Depending on the mean stellar age and star formation history, the restframe
equivalent width of the H$\beta$ line {\it in absorption} may be in the range
$\sim 5 - 10$ \AA.  Our observed equivalent width of the H$\beta$ line in 
emission is 48 \AA, corresponding to $\sim 24.5$ \AA\ in the restframe.  Thus
the correction to the derived Balmer emission line luminosities may be as 
large as 20\%, or higher.  Correspondingly, our derived SFR should be adjusted
upwards by such a factor.

Finally, an alternative estimate of the SFR can be obtained from the continuum
luminosity at $\lambda_{rest} = 2800$\AA\ (\cite{Mad98}). 
The observed continuum flux at the corresponding wavelength of 5505 \AA\ in
our July 7 spectrum is $\sim 3.1 ~\mu$Jy.  Correcting for the Galactic
foreground extinction, the value becomes $\sim 3.7 ~\mu$Jy.  We estimate that
the OT contribution at this time was about 30\% of the total flux, 
and thus the flux from the galaxy itself to be about $2.6 ~\mu$Jy.
The implied restframe power at this wavelength is then
$P_{2800} \approx 6.13 \times 10^{28}$ erg s$^{-1}$ Hz$^{-1}$.
Using the \cite{Mad98} estimator, the corresponding star formation rate is
$SFR \approx 7.8 ~M_\odot ~{\rm yr}^{-1}$, 
again without any restframe extinction correction, and in an excellent
agreement with the estimate derived from the H$\beta$ line.  Adopting restframe
$A_V \approx 0.3$ mag would increase these numbers by 70\%, i.e., to 
$SFR \approx 13 ~M_\odot ~{\rm yr}^{-1}$.

We thus conclude that the conservative lower limit to the star formation rate is
$SFR > 7 ~M_\odot ~{\rm yr}^{-1}$ 
(with no restframe extinction correction at all), at more realistically 
$SFR \approx 10 ~M_\odot ~{\rm yr}^{-1}$ or even higher, depending on the
actual extinction value.

We note that these estimates exclude any star formation which may be heavily
obscured, and whose diagnostics would thus escape our detection.  In this case 
one may expect to detect the dust heated by the newly formed stars
(\cite{hug98}), or else the measure the non-thermal radiation, produced as a
by-product of the on-going star formation (\cite{con92}).  Unless our $SFR$ is
underestimated by an order of magnitude or more, we determined that both
processes would be expected at a level of $\sim 10 \mu$Jy, and thus cannot be
easily detected with current telescopes. 

This star formation rate is higher than in either one of the two GRB host
galaxies with previously determined redshifts (GRB 970508 and GRB 971214),
but is still not exceptional.  This finding gives some modest support to the
idea that GRBs may be closely related to recent star formation
(e.g., \cite{tot97}, \cite{wij98}).

\acknowledgments

We are grateful to the Directors of the W. M. Keck and Palomar observatories,
Drs. F. Chaffee and W. Sargent, for the allocation of service observing time
for the initial observations reported here, and to the staff of WMKO for their
expert assistance with the observing runs, and to the entire BeppoSAX team for
their efforts. 
This work was supported in part by the Bressler Foundation (SGD), and grants 
from the NSF and NASA (SRK).

\clearpage

\begin{deluxetable}{lllll}
\tablecaption{Observed Emission Lines 
\label{tab1}
}
\tablenum{1}
\tablewidth{0pt}
\tablehead{
\colhead{$\lambda_{obs}$ (\AA)}
&\colhead{Line ID}
&\colhead{$z_{em}$}
&\colhead{$f_{obs}$\tablenotemark{a}} 
&\colhead{$f_{corr}$\tablenotemark{b}} 
}
\startdata
7327.9 & [O II]    3727.4 & 0.9660 & 26.9 & 30.4 \nl
8065   & H$\delta$ 4101.7 & 0.9663 & 1.3  & 1.45 \nl
8533.5 & H$\gamma$ 4340.5 & 0.9661 & 2.55 & 2.8  \nl
9558.1 & H$\beta$  4861.3 & 0.9662 & 5.65 & 6.1  \nl
9751   & [O III]   4958.9 & 0.9664 & 2.5  & 2.7  \nl
\enddata
\tablenotetext{a}{
Observed fluxes in units of $10^{-17}$ erg cm$^{-2}$ s$^{-1}$.
}
\tablenotetext{b}{
Corrected for the Galactic extinction.
}
\end{deluxetable}


\begin{deluxetable}{llll}
\tablecaption{Observed Absorption Lines 
\label{tab2}
}
\tablenum{2}
\tablewidth{0pt}
\tablehead{
\colhead{$\lambda_{obs}$ (\AA)}
&\colhead{$\lambda_{vac}$ (\AA)}
&\colhead{Line ID}
&\colhead{$z_{abs}$}
}
\startdata
4605.9 & 4607.2 & Fe II 2344.2 & 0.9653 \nl
4669.9 & 4671.2 & Fe II 2374.5 & 0.9672 \nl
4683.6 & 4684.9 & Fe II 2382.8 & 0.9662 \nl
5083.8 & 5085.2 & Fe II 2586.7 & 0.9654 \nl
5109.9 & 5111.3 & Fe II 2600.2 & 0.9657 \nl
5493.6 & 5495.2 & Mg II 2796.4 & 0.9651 \nl
5508.1 & 5509.6 & Mg II 2803.5 & 0.9653 \nl
5604.6 & 5606.2 & Mg I  2852.1 & 0.9656 \nl
\enddata
\end{deluxetable}

\clearpage

\noindent\bf{Fig. 1.}~~
The initial spectrum of the host galaxy of GRB 980703, obtained at the Keck
telescope on 7 July 1998 UT, when the OT was still contributing about one third
of the observed continuum.  The spectrum was smoothed with a gaussian with a 
$\sigma = 5$\AA, roughly corresponding to the instrumental resolution. 
Prominent emission and absorption lines are labeled.

\bigskip
\bigskip

\noindent\bf{Fig. 2.}~~
The far-red spectrum of the host galaxy of GRB 980703, obtained at the Keck
telescope on 19 July 1998 UT.  The spectrum was smoothed with a gaussian with a 
$\sigma = 2$\AA, roughly corresponding to the instrumental resolution. 
Prominent emission lines are labeled.  The contribution of the OT to the
spectrum was negligible at this point.

\bigskip
\bigskip

\noindent\bf{Fig. 3.}~~
A zoom-in on the spectrum from July 19, with the prominent absorption 
features labeled.  They are most likely caused by the ISM in the GRB host
galaxy itself.  The spectrum was smoothed with a gaussian with a 
$\sigma = 1$ pixel $\approx 2.5$\AA.

\bigskip
\bigskip

\begin{figure}
\figurenum{1}
\plotone{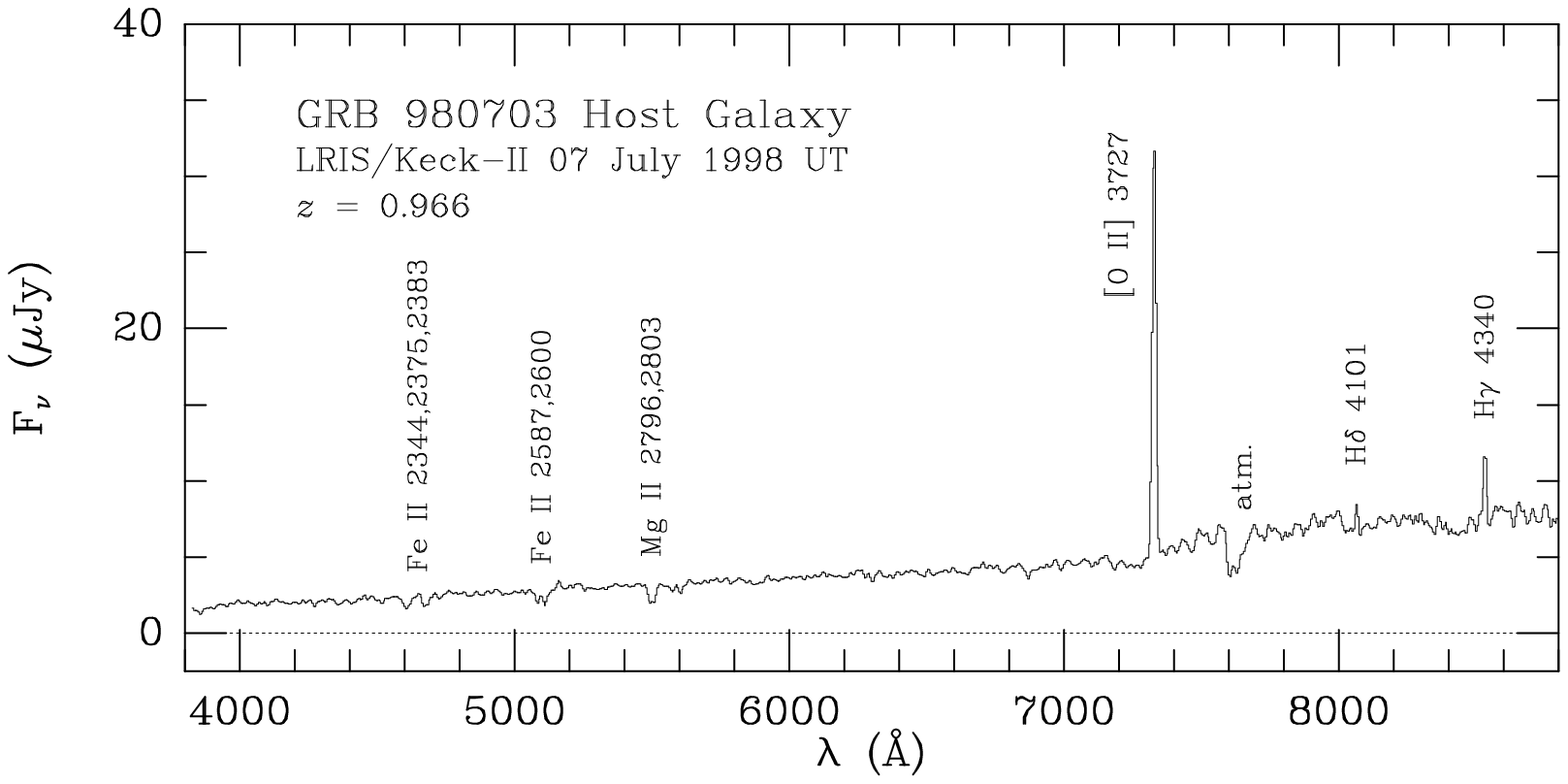}
\epsscale{1.0}
\caption{
\label{fig1}
}
\end{figure}

\clearpage

\begin{figure}
\figurenum{2}
\plotone{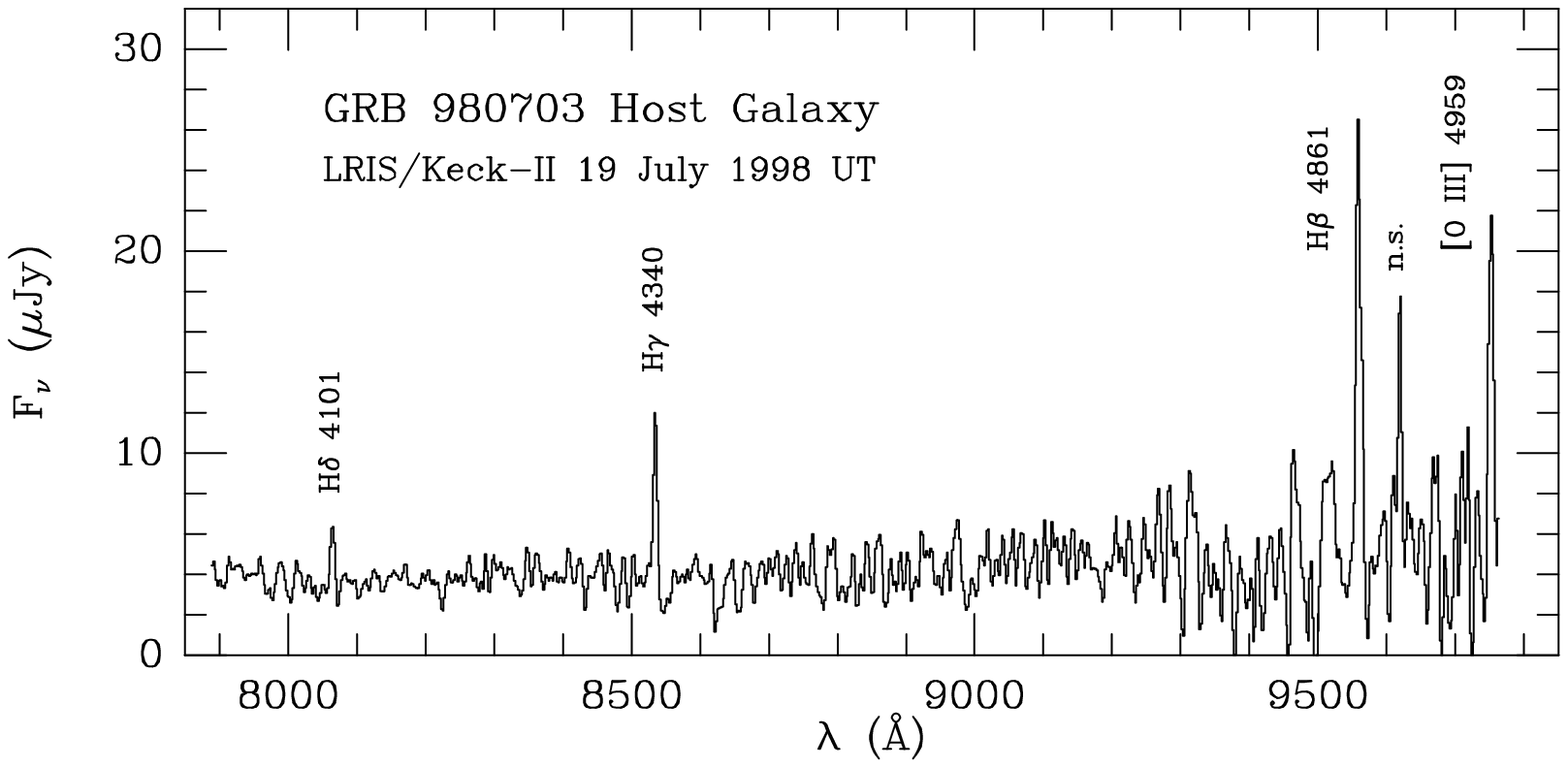}
\epsscale{1.0}
\caption{
\label{fig2}
}
\end{figure}

\clearpage

\begin{figure}
\figurenum{3}
\plotone{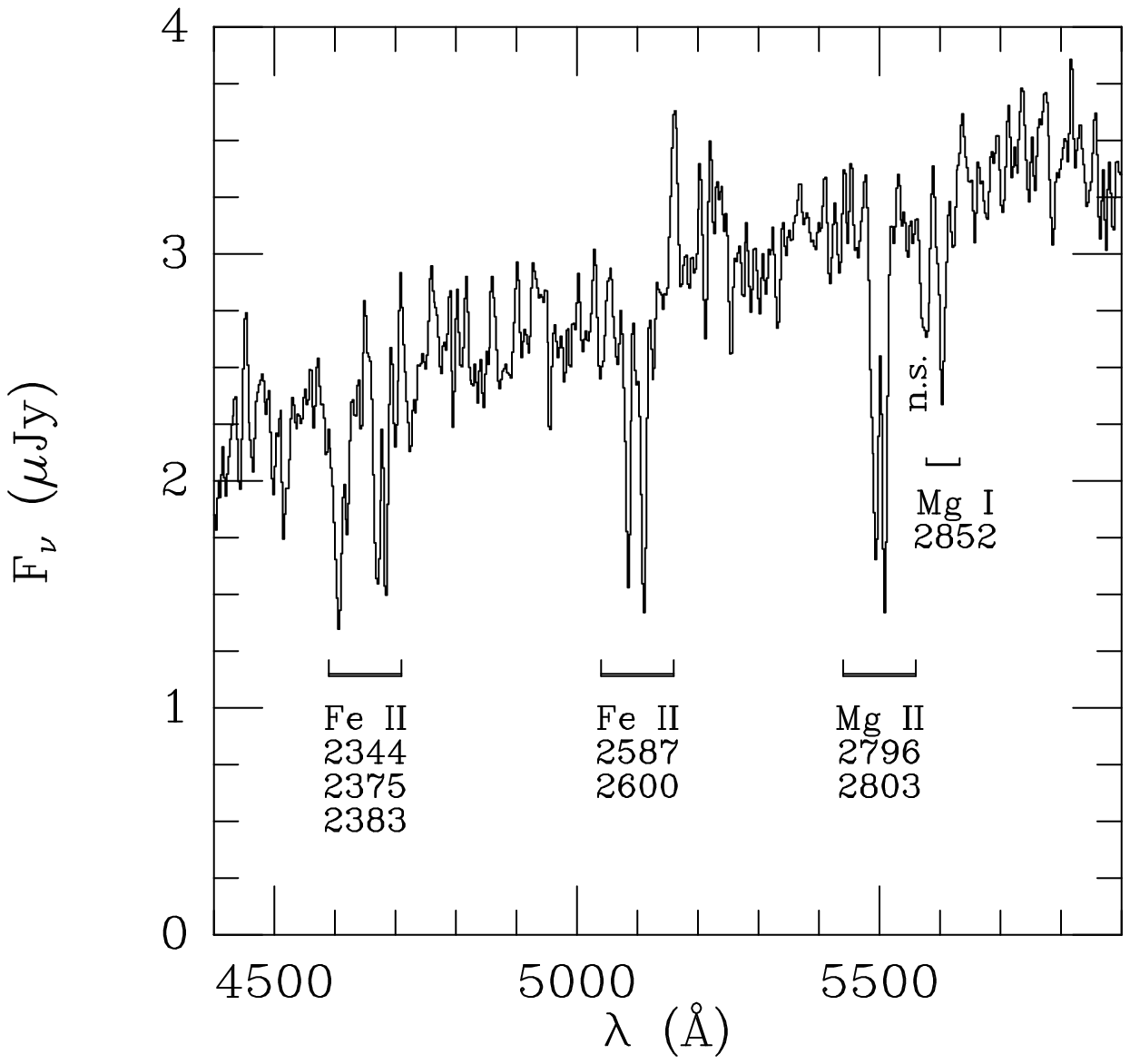}
\epsscale{1.0}
\caption{
\label{fig3}
}
\end{figure}


\end{document}